\documentclass[journal,onecolumn,11pt]{IEEEtran}
\usepackage{cite}
\usepackage{amsmath,amssymb,amsfonts}
\usepackage{algorithmic}
\usepackage{graphicx}
\usepackage{textcomp}
\usepackage[dvipsnames]{xcolor}
    
\usepackage[hidelinks]{hyperref}
\usepackage{mathtools}
\usepackage{bbm}
\usepackage{algorithmic, algorithm}
\usepackage{etoolbox}
\AtBeginEnvironment{bmatrix}{\setlength{\arraycolsep}{2pt}}

\usepackage{glossaries}
\glsdisablehyper
\newacronym{3gpp}{3GPP}{3rd Generation Partnership Project}
\newacronym{bs}{BS}{base station}
\newacronym{ue}{UE}{user equipment}
\newacronym{mimo}{MIMO}{multiple-input multiple-output}
\newacronym{mdt}{MDT}{minimization of drive tests}
\newacronym{urllc}{URLLC}{ultra-reliable low-latency communication}
\newacronym{rss}{RSS}{received signal strength}
\newacronym{rssi}{RSSI}{received signal strength indicator}
\newacronym[longplural={Gaussian processes }]{gp}{GP}{Gaussian processes}
\newacronym{csi}{CSI}{channel state information}
\newacronym{iot}{IoT}{internet of things}
\newacronym{pdf}{PDF}{probability density function}
\newacronym{cdf}{CDF}{cumulative distribution function}
\newacronym{snr}{SNR}{signal-to-noise ratio}
\newacronym{sinr}{SINR}{signal-to-interference-plus-noise ratio}
\newacronym{los}{LOS}{line of sight}
\newacronym{nlos}{NLOS}{non-line of sight}
\newacronym{tdma}{TDMA}{time-division multiple access}
\newacronym{3d}{3D}{three-dimensional}
\newacronym{pcr}{PCR}{probably correct reliability}

\begin{document}

\title{Delivering Ultra-Reliable Low-Latency Communications via Statistical Radio Maps}

\author{Tobias~Kallehauge, Anders~E.~Kalør, Pablo~Ramírez-Espinosa, Maxime~Guillaud, and Petar~Popovski
\thanks{T. Kallehauge, A. E. Kalør, P. Ramírez-Espinosa and P. Popovski are with the Department of Electronic Systems, Aalborg University, Aalborg 9220, Denmark (e-mail: \{tkal, aek, pres, petarp\}@es.aau.dk). M. Guillaud is with Huawei Technologies, Paris, France (e-mail: maxime.guillaud@huawei.com).}}%

\maketitle

\begin{abstract}
High reliability guarantees for Ultra-Reliable Low-Latency Communications (URLLC) require accurate knowledge of channel statistics, used as an input for rate selection. Exploiting the spatial consistency of channel statistics arises as a promising solution, allowing a base station to predict the propagation conditions and select the communication parameters for a new user from samples collected from previous users of the network. Based on this idea, this article provides a timely framework to exploit long-range channel spatial correlation through so-called statistical radio maps, enabling URLLC communications with given statistical guarantees. The framework is exemplified by predicting the channel capacity distribution both in a location-based radio map and in a latent space rendered by a channel chart, the latter being a localization-free approach based on channel state information (CSI). It is also shown how to use the maps to select the transmission rate in a new location that achieves a target level of reliability. Finally, several future directions and research challenges are also discussed.  
\end{abstract}

\section{Introduction}

\Gls{urllc} is undoubtedly one of the most significant and challenging novelty introduced in 5G. Its stringent requirements in terms of latency and reliability are envisioned to enable a plethora of new use cases, ranging from medical applications to industrial automation. Yet, the most extreme requirements, such as reliabilities in the order of $1-10^{-8}$ as required for instance by motion control applications~\cite{ts22.104}, are still far from being met, and further developments are needed as the industry move towards 6G. Traditionally, high reliability has been achieved by using high levels of diversity. Diversity in time, manifested through coding and retransmissions, translates into higher latency, unacceptable for many \gls{urllc} applications. Increasing bandwidth to attain frequency diversity arises as a solution to meet reliability and latency constraints \cite{Popovski2019}, but the frequency spectrum is a scarce resource, and the available bandwidth is limited in practical scenarios. 

One of the central challenges in providing \gls{urllc} is the inherent randomness of the wireless channel. Instantaneous \gls{csi} acquisition is thus ubiquitous in any communication protocol, where the transmission parameters, such as resource allocation, power, and rate, are selected based on the estimated channel information. However, the ultra-reliable regime challenges this \gls{csi}-based paradigm in different ways:
\begin{enumerate}
    \item Estimating the channel introduces a non-negligible latency that may not be acceptable for mission-critical applications. This latency comes not only from the pilot transmission but also from the processing --- for example, the most commonly used estimators in \gls{mimo} systems require matrix inversions.
    \item Once the channel is estimated, it is assumed to be constant during the whole transmission block, which may be too strong an assumption when dealing with extremely low error probabilities \cite{Swamy2019}.
    \item Besides instantaneous CSI, knowledge of channel statistics is essential to be able to offer reliability guarantees that span over multiple transmissions. 
    \item Channel estimation is not error-free (due to pilot contamination, noise, interference, etc.), and this error is propagated into the selected transmission parameters. The error may be acceptable when working with error probabilities around $10^{-2}$, but when targeting probabilities of $10^{-5}$ or lower, even a small estimation error can cause the requirements to be violated. 
\end{enumerate}

An alternative to estimating the instantaneous \gls{csi} is to rely on \textit{statistical knowledge} of the channel by configuring the system to achieve target reliability with a specified confidence level \cite{Angjelichinoski2019}. For instance, given an estimate of the channel statistics, the transmission parameters can be selected such that communication is successful with a controlled likelihood. This approach circumvents the issues of estimating instantaneous \gls{csi} but comes at the price of (offline) collecting  enough samples to estimate the channel statistics. Moreover, in some cases the required number of samples may be extremely large --- even prohibitive --- especially when non-parametric estimation is carried out \cite{Angjelichinoski2019}. In essence, we would like our systems to operate without errors for longer than we can afford to spend collecting channel samples. Although the number of samples required for estimation can be reduced by using parametric models, these introduce a risk of model mismatch. This can be particularly severe, considering that most of the channel models available in the literature are not designed for \gls{urllc} and may not accurately characterize rare events, which can have significant impact in the \gls{urllc} regime.

Reducing the required number of samples to get statistical knowledge of the channel without compromising the statistical guarantees arises then as a necessary but challenging goal. One promising solution is to exploit observations from \emph{all} users in the network and use this knowledge to select the transmission parameters for users in similar conditions. Generalizing this idea, we ask \textit{how to learn from past experiences in similar wireless communication scenarios and how to use that knowledge to predict the performance in a new one?} Aiming to answer this question, in this article we illustrate how data collected across a network can be used to select transmission parameters for new users by leveraging the spatial consistency inherent to any natural propagation environment, while providing statistical guarantees required in the context of \gls{urllc}.

\section{The Challenge of Estimating Channel Statistics for URLLC}
Before going further into the topic of leveraging past observations using spatial correlation of channel statistics, we illustrate the inherent difficulties of estimating channel statistics for \gls{urllc} through an example. 

\begin{figure}
    \centering
    \includegraphics[width=\linewidth]{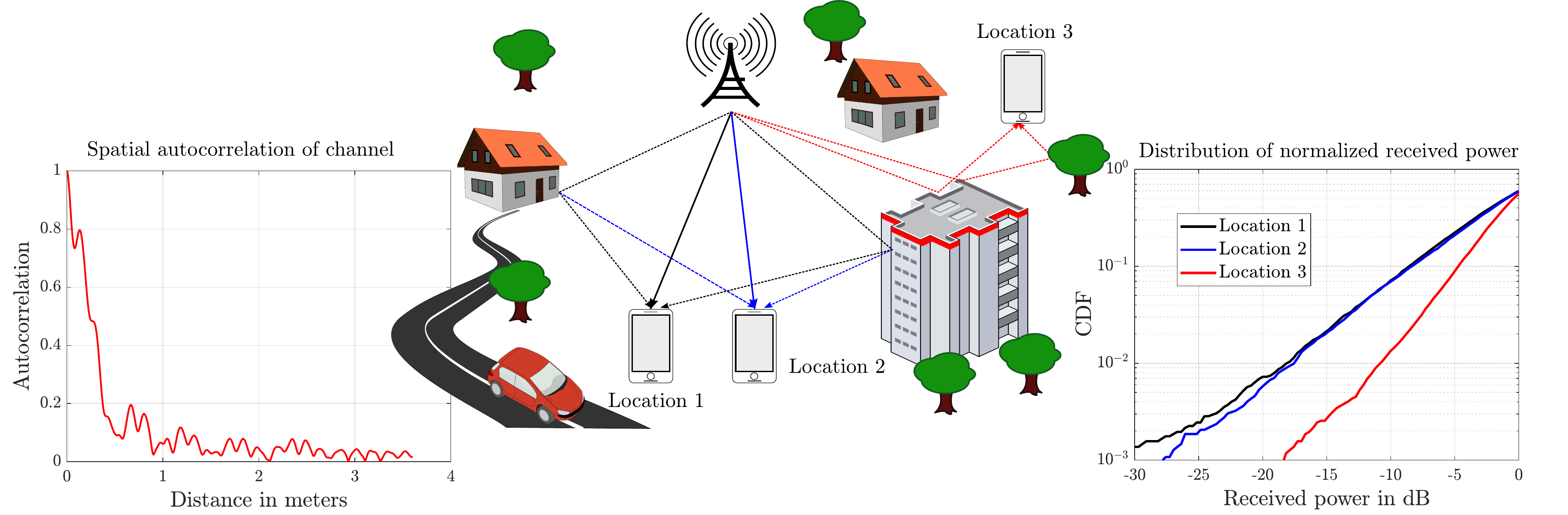}
    \caption{Illustration of urban propagation environment and incoming paths, where the autocorrelation function of the received narrowband channel is plotted as a user moves from locations 1 to 2. Also, the statistics of the received power at three different locations are plotted in the form of their \glspl{cdf}. Location 1 is at $(-20, -60)$ m, location 2 is at $(-12, -56)$ m, location 3 is at $(30, 40)$ m, and the base station is at $(0,0)$. Data is obtained using QuaDRiGa \cite{Quadriga} and does not necessarily match the illustrated environment.}\label{fig:propagation}
\end{figure}

\begin{figure}
    \centering
    \includegraphics[width = .8\linewidth]{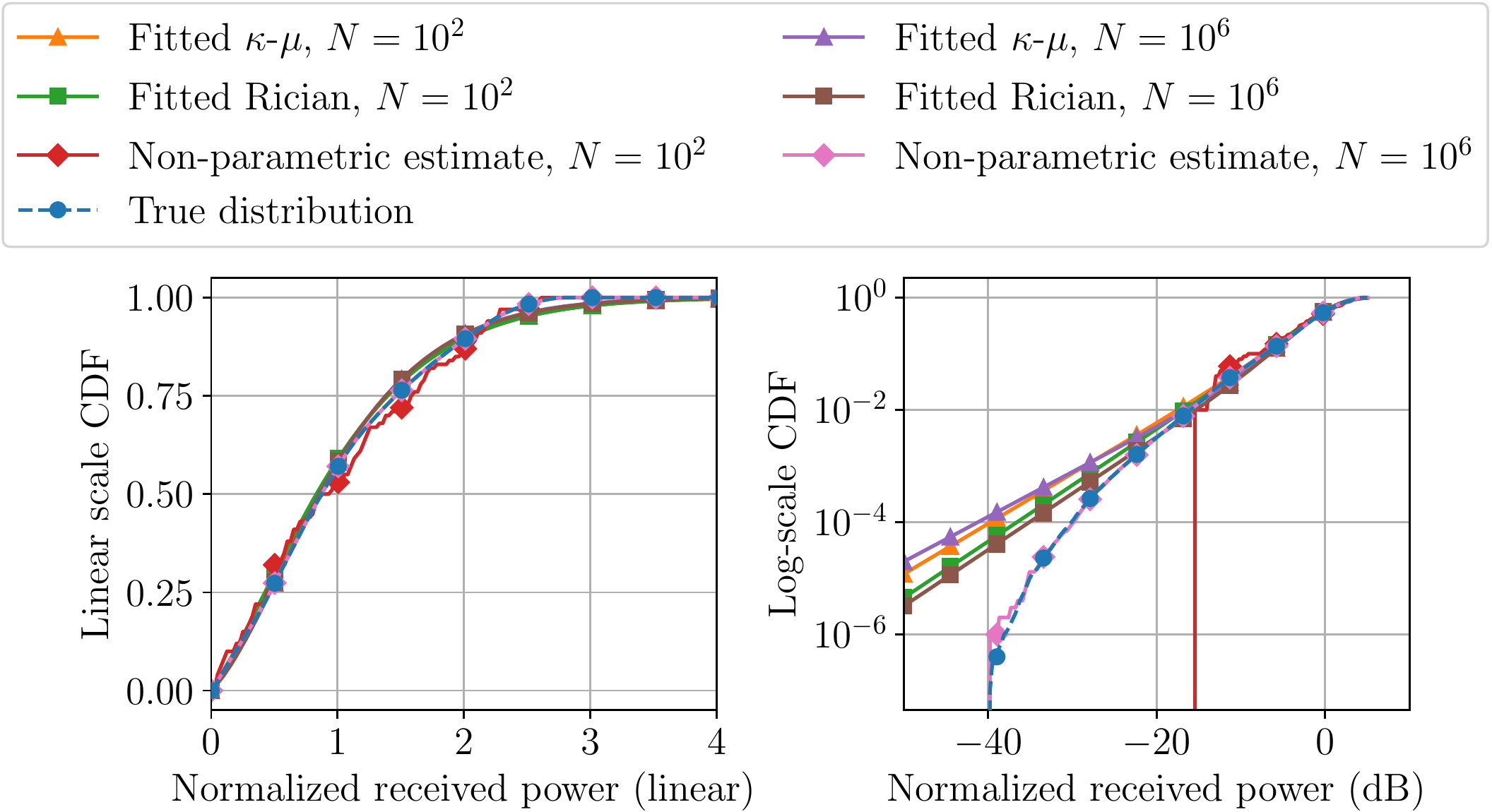}
    \caption{Fitted \glspl{cdf} to normalized received power samples. The true distribution is obtained as the sum of $7$ paths with different amplitudes and uniformly distributed phases. The resulting \gls{cdf} is fitted through maximum likelihood to a $\kappa-\mu$ distribution, a Rician distribution, and a non-parametric estimator, each with $N$ number of samples. Both linear and logarithmic scales are depicted.}
    \label{fig:cdf_fit}
\end{figure}

Consider a standard cellular system with a transmitter (e.g., a \gls{bs}) and a receiver (some user). Due to the wireless propagation and its dynamics, the received power is well characterized as a random variable with certain statistics imposed by the propagation environment and therefore depend on the location, as illustrated in Fig. \ref{fig:propagation}. As mentioned before, characterizing these statistics arises as a necessity for achieving reliable communication. However, due to the stringent requirements of \gls{urllc}, the system is faced with the challenge of estimating the \emph{tail} of the received power distribution, reflecting rare events, without collecting an excessive number of channel samples. These challenges are illustrated in Fig. \ref{fig:cdf_fit}, where the actual \gls{cdf} of the normalized received power is fitted using a different number of samples. The power samples are obtained as the sum of multiple paths with different amplitudes and random phases, which models the signal reflections. Maximum likelihood estimation is then used to fit the received samples to some widely-used parametric models together with a non-parametric estimator. Judging from the linear scale result, it may seem that the model mismatch effect is negligible. However, when looking at the tail --- as required in \gls{urllc} --- in the logarithmic representation, the impact of model mismatch is clear: regardless of the number of samples available, the parametric models cannot fit the true channel statistics. Even though the incurred error is small for, e.g., probabilities around $10^{-2}$ (the regime where many communication systems operate), it is completely unacceptable for lower probabilities like $10^{-5}$ that \gls{urllc} systems target. The non-parametric estimator naturally converges to the true statistics, but as mentioned before, it requires a very large number of samples, which is often prohibitive in practice. 

To prevent model mismatch, in this article, we work with non-parametric estimators, where the problem of the sample size is alleviated by the fact that the past observations used for predicting channel statistics for a new user can be gathered over a long period and therefore do not affect the latency induced by selecting the transmission rate. 

\section{Spatial Correlation of Wireless Channel Distributions and Radio Maps}
In any wireless communications link, the transmitted radio signal is subject to randomizing effects from reflection, diffraction, and scattering as it travels from
transmitter to receiver, giving rise to multipath propagation that alters the phase and amplitude of the received radio signals, as illustrated in Fig. \ref{fig:propagation}. These changes are typically modeled under the umbrella of three major effects: pathloss, shadowing, and fast fading \cite{Goldsmith2005}. Pathloss and shadowing refer to the attenuation the signal experiences due to the distance and the presence of blockages and large objects. Naturally this attenuation is correlated over space as a rule of thumb on scales of $100$--$1000$ m and $10$--$100$ m, respectively. This smooth variation is captured in conventional radio maps, depicting the average received power over space, see e.g.,~\cite{Chowdappa2018, Wang2020} and references therein. However, average values are not enough for \gls{urllc} since they provide little information about the actual channel dynamics and statistics; instead, characterizing the fast fading is essential to provide statistical guarantees. 

Fast fading, or simply fading, arises as the constructive and destructive sum of the multiple unresolvable paths arriving at the receiver and therefore occurs at a much smaller scale than shadowing and pathloss. In fact, fast fading varies in the order of only a few wavelengths as it depends on the phases of the different paths, each of which can change rapidly with only a slight movement of the transmitter, receiver, or elements in the environment. 
Predicting the instantaneous values for small-scale fading based on spatial correlation is, therefore, infeasible, which is why radio maps tend to model only path loss and shadowing by averaging out fast fading. However, the conclusion that small-scale fading cannot be spatially predicted only applies to the instantaneous values and not its statistics, in the sense that the probability distributions of fast fading for two nearby locations are likely to be similar. 
The spatial behavior of fast fading is illustrated in Fig. \ref{fig:propagation}, where the spatial autocorrelation of the channel and the received power distribution are plotted for a typical urban environment. We can see that, even though the autocorrelation of the channel vanishes rapidly as the user moves due to the changes in the phases of the arriving paths, the distributions of the signal power received by nearby points are similar.
This leads to the concept of \textit{statistical radio maps} \cite{Kallehauge2022_globecom}, which, in striking contrast to conventional radio maps, aim to exploit the \textit{spatial consistency} of the channel by establishing a relationship between location and channel statistics beyond average values. 

To illustrate the use of statistical radio maps, consider the same cellular system as in Fig. \ref{fig:propagation}. As a new user joins the network, the \gls{bs} can use samples collected from previous users to build a radio map and \textit{infer} the channel statistics that new users may experience based on the location of the new user. From a \gls{urllc} viewpoint, this procedure can significantly reduce the number of samples required for the new user to estimate relevant channel statistics (beyond simply the average value) by predicting these statistics directly on the map, thereby reducing the latency. As the previous discussion suggests, using geographic location as the metric for building statistical radio maps appears as a natural choice since spatial correlation is inherent in any real propagation environment. The location, however, only serves as a proxy for characterizing the underlying channel, and other variables can serve in the place of location. Indeed, we demonstrate how the relation between high-dimensional \gls{csi} measurements and long-term channel statistics can be learned through the use of a \textit{channel chart} \cite{Studer2018} --- a ``map-like'' latent space of lower dimension than the \gls{csi} that is not based on geographic locations.

Note that although we motivated the use of statistical radio maps to model the received signal power, they can be used to capture the statistics of other, equally relevant, variables. In general, they may characterize any value that can be useful to select the transmission parameters, such as \gls{snr} or \gls{sinr}. However, for clarity, we limit our focus to the channel capacity as the value of interest in the remainder of the article. Specifically, we aim to characterize the \textit{$\epsilon$-outage capacity}, denoted $C_{\epsilon}$, and defined as the transmission rate (or simply rate) supported by the channel with probability at least $1 - \epsilon$ for some specified maximum desired failure probability $\epsilon$.

\section{Data-Driven Construction of Statistical Radio Maps}
This section presents a general approach to constructing statistical radio maps from empirical channel observations ---  specific examples are given in the next section. We first explain how a map can be constructed for geographic locations, and then show how the geographic locations can be replaced by a channel chart. As mentioned, we focus on the case where the maps are used to select the transmission rate for new users, and we show how to incorporate the uncertainty of the predicted maps to achieve statistical reliability guarantees. 

For both the localization and chart-based approach, we assume the availability of a training dataset, $\mathcal{D}$, comprising a large number of --- noiseless --- received power samples from previous transmissions between a central node (e.g., \gls{bs}) and users at different locations within an area. Following the previous discussions, and by assuming the noise power is constant and known, the outage capacity $C_{\epsilon}$ is estimated at each location from the received power samples non-parametrically.
The dataset additionally contains either the location or a \gls{csi} measurement for each user among the previous transmissions.

\subsection{Location-based Maps}
\begin{figure}
    \centering
    \includegraphics[width=\linewidth]{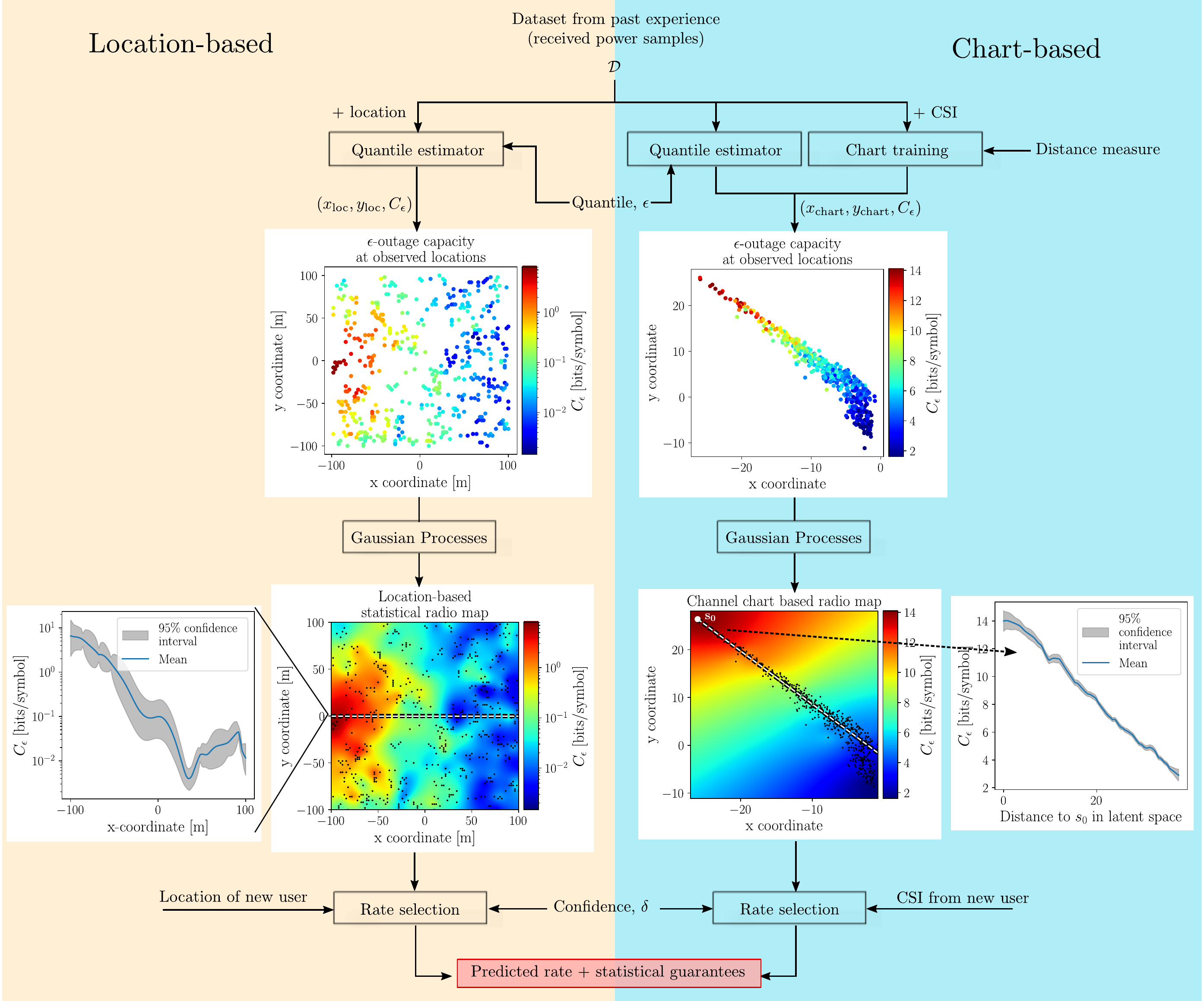}
    \caption{Diagram of the two data-driven methods to construct statistical radio maps introduced in the article.}
    \label{fig:diagram_map_construction}
\end{figure}

The process of constructing a map is straightforward when the locations are known, e.g., obtained using GPS. First, the $\epsilon$-outage capacities are estimated non-parametrically at each observed location from the received power samples for a given choice of $\epsilon$ between $0$ and $1$. This is done similarly as in \cite{Kallehauge2022_globecom} by computing the channel capacity for each received power sample, and then selecting the capacity as the $\epsilon$-quantile of the empirical distribution. This yields an ``incomplete'' map where the outage capacities are known only at the locations contained in the dataset, as shown on the upper left plot in Fig.~\ref{fig:diagram_map_construction}. We then complete the map by spatial interpolation, which can be done using e.g., \textit{\glspl{gp}} \cite{Rasmussen2006} and \textit{Kriging interpolation} \cite[ch. 3]{chiles2012geostatistics}. For the construction of the statistical radio map, we use a \gls{gp} as it allows us to obtain the posterior distribution of its predictions, which is central for providing statistical guarantees for \gls{urllc}.

Interpolating using \glspl{gp} first involves fitting a spatial model for the mean and covariance between points on the map and then obtaining predictions at unobserved locations by assuming that all points on the map follow a multivariate Gaussian distribution with parameters from the fitted model --- the complete procedure is described in \cite{Kallehauge2022_globecom}. The result of the interpolation method is a statistical radio map, where the predictive distribution of the outage capacity can be computed for each location. The mean of the predictive distribution can be considered the predicted quantile (as shown in the lower left plots of Fig.~\ref{fig:diagram_map_construction}), and the variance quantifies the uncertainty in the predicted value resulting from both observation noise and the variance across the spatial domain (as shown with the $95\%$ confidence interval in Fig.~\ref{fig:diagram_map_construction}).

Using the predictive distribution, we select the transmission rate conservatively as the $\delta$-quantile of the predicted outage capacity for some low value $\delta$, i.e., the rate is selected relatively lower than the predicted mean controlled by $\delta$. This choice of rate is motivated by the concept of \emph{meta-probabilities}, which has previously been exploited for \gls{urllc}~\cite{Angjelichinoski2019,Kallehauge2022_globecom}, and which takes into account both the random sampling process and the randomness in the spatial distribution of the outage capacities. Provided that the chosen model matches the observations, $\delta$ has the intuitive meaning of the probability that the selected rate is greater than the $\epsilon$-outage capacity when taking into account the spatial distribution, and thus the probability of violating the reliability constraint characterized by $\epsilon$. Note that this differs from traditional point estimators, such as maximum likelihood or maximum a posteriori, which predict the most probable outage capacity and are likely to predict a rate that violates the reliability constraint approximately 50\% of the time.

\subsection{Chart-based Maps} \label{subsec:ch_chart}
The radio map presented in the previous section requires that the location of all the users in the scenario is known, which may not be available in practice, e.g., in indoor environments. The location-based approach also adds additional latency caused by the transmission of the location to the \gls{bs}, which may not be acceptable for \gls{urllc}, especially in scenarios with mobility. An appealing idea to solve these issues, particularly suitable for massive MIMO scenarios, is to eliminate the need for location information by instead extracting a snapshot of the physical environment, such as the various paths, angle of arrivals, etc., from the \gls{csi}. Note that such an approach would be different from obtaining instantaneous \gls{csi} prior to each transmission, as it only needs to be updated when the dominant paths change significantly.
Constructing the map directly in the \gls{csi} space would then allow the outage capacity to be predicted for previously unseen \gls{csi} values similar to the location-based approach while only relying on the transmission of pilot symbols by the users.

However, applying \gls{gp} regression directly in the \gls{csi} space is impractical due to the curse of dimensionality and the high complexity of employing \glspl{gp} in high dimensions.
Instead, a more promising way to exploit the \gls{csi} is through the use of  \emph{channel charting}~\cite{Studer2018, Ferrand2021}, where the goal is to learn a mapping (e.g., through a neural network) from the high-dimensional \gls{csi} space to a low-dimensional latent space. The mapping is learned using some desired distance measure so that points that are close according to the chosen measure are also close in the latent space. While the latent space is often designed to resemble the geographic space by using physical distance metrics, more suitable distance measures in the context of predicting the outage capacity could be chosen. For instance, some metric between the probabilistic distribution of the instantaneous channel capacity would promote a channel chart that is spatially consistent with statistics of the capacity (e.g., the outage capacity).

Once a suitable mapping from the \gls{csi} space to a latent space has been obtained, \gls{gp} regression can be applied in the latent space precisely as done for the location-based map, and the resulting map can similarly be used to select a transmission rate. This procedure is illustrated in the right half of Fig.~\ref{fig:diagram_map_construction}. 
Note that the points describing the channel statistics to predict (such as the outage capacity) do not have to be from the same radio interface as the one used to transmit the pilots for the \gls{csi} as long as the \gls{csi} is correlated with the channel statistics. For instance, if a \gls{bs} operates several different frequency bands, channel charts can be created for each band while using only \gls{csi} measurements from one band. This allows, for instance, a system to exploit a massive MIMO antenna operating at one frequency band to obtain a rich \gls{csi} snapshot that can be used to predict channel statistics for a \gls{urllc} application operating at a different frequency band.

\section{Case Study} \label{sec:case_study}
In this section, we illustrate the use of statistical radio maps through two \gls{urllc} scenarios. The first scenario shows location-based maps, whereas the second scenario demonstrates chart-based maps. Both scenarios comprise a single \gls{bs} located at $(-100,0,10)$ m that serves several users within a square cell of $200\times 200$ m$^2$. The users' locations within the cell are assumed to be drawn from a clustered Thomas point process, which mimics, e.g., mobile users taking measurements along a busy street in an urban environment, and that all users have the same height of $1.5$ meters. The channels between the \gls{bs} and the users are simulated in an urban micro-cell non-line-of-sight scenario using the QuaDRiGa simulator~\cite[p. 81]{Quadriga} --- further details are given for each scenario. Our goal is to use a statistical radio map to select the rate of a new user at a random location within the cell, such that the outage probability experienced by the user is below $\epsilon=10^{-3}$ with confidence $\delta = 10^{-3}$.

\subsection{Location-based rate selection}
\begin{figure}
    \centering
    \includegraphics[width = .8\linewidth]{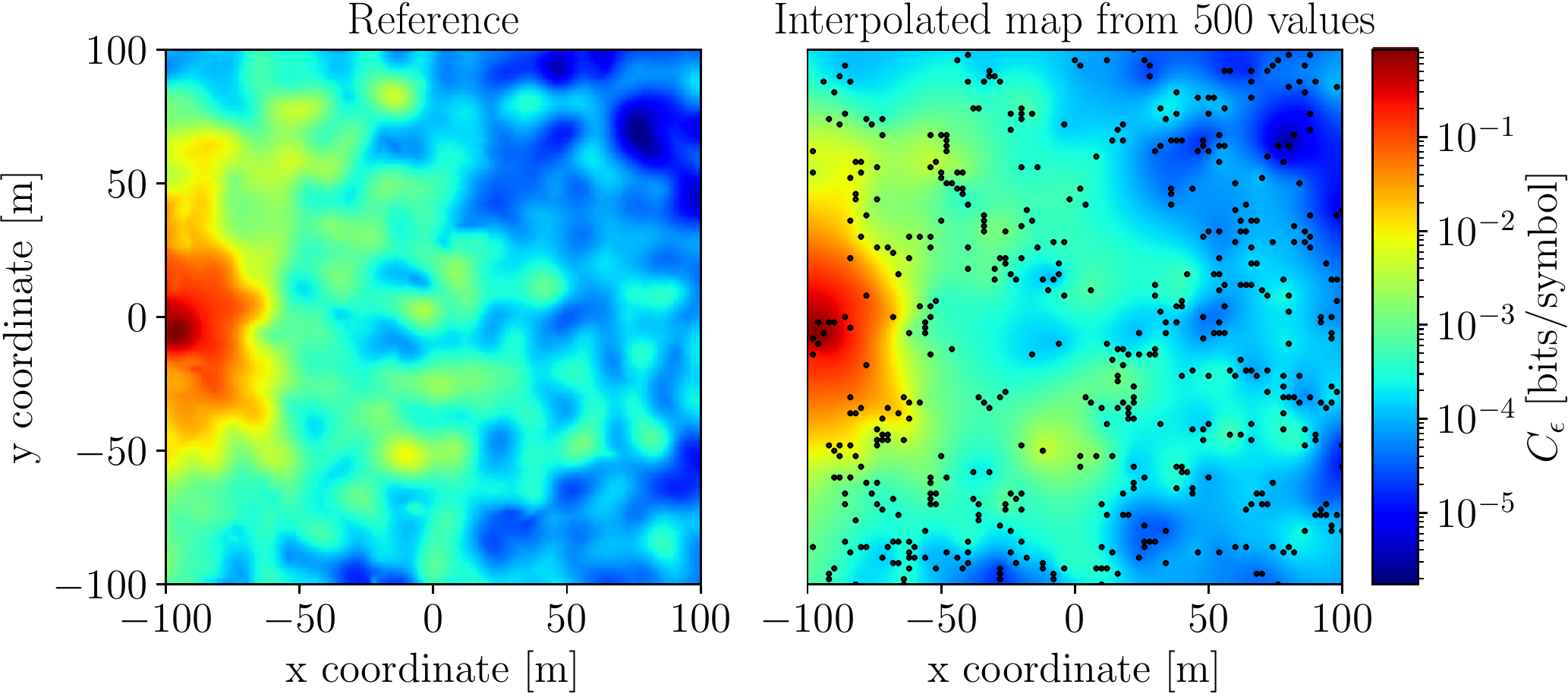}
    \caption{Location-based statistical radio map. Left: True outage capacity. Right: Maximum a posteriori estimates of the outage capacities obtained by interpolating between empirical outage capacities measured at the black dots.}
    \label{fig:interpolartion_location}
\end{figure}

\begin{figure}
    \centering
    \includegraphics[width = .5\linewidth]{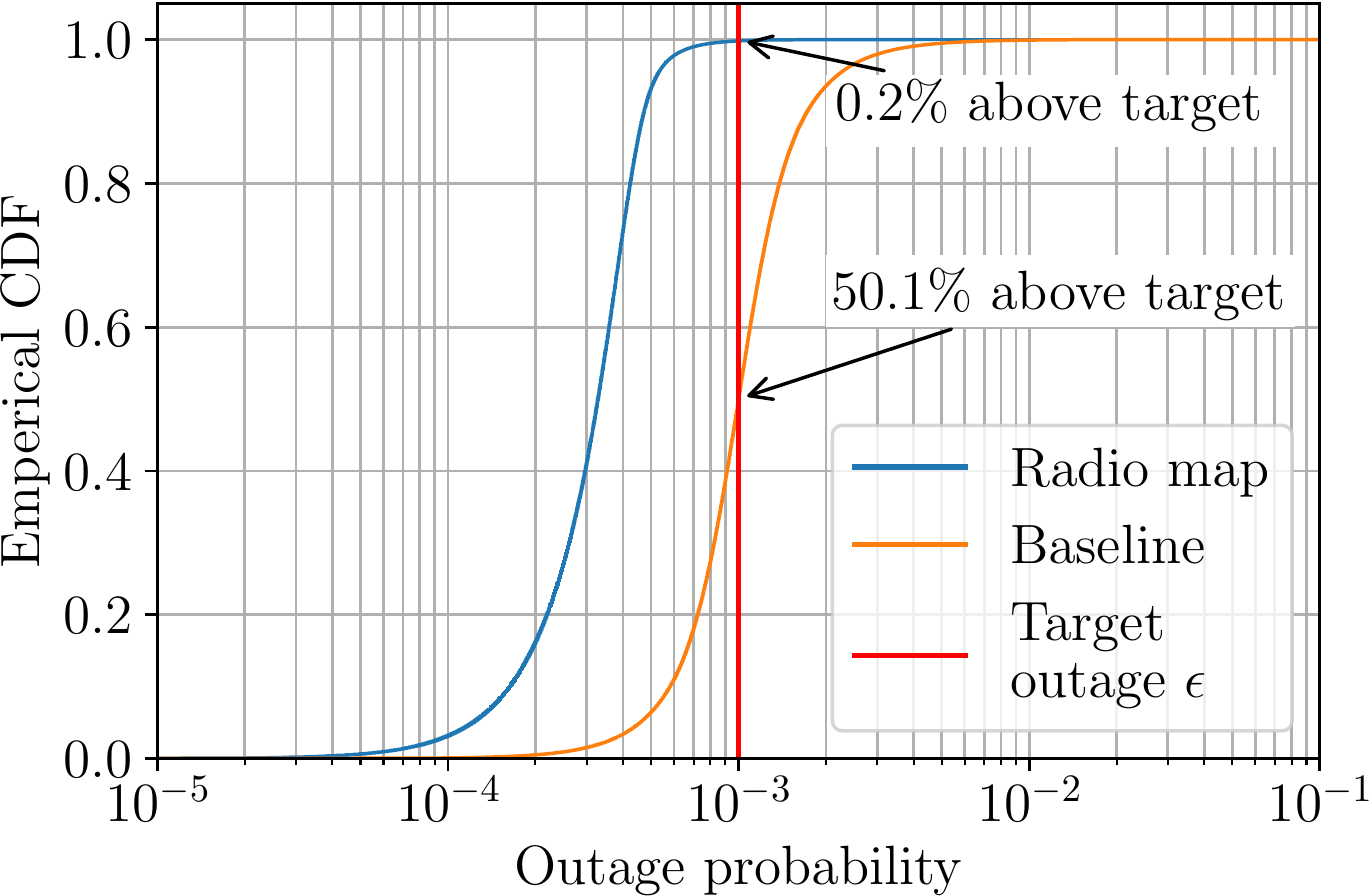}
    \caption{Performance of rate selection using the location-based radio map with $\epsilon = 10^{-3}$ and $\delta=10^{-3}$. Only $0.2\%$ of the rates obtained using the radio map violate the reliability requirements, whereas $50.1\%$ of the rates obtained with the baseline scheme are above the outage capacity.}
    \label{fig:performance_location}
\end{figure}

A SISO narrowband channel in the 800~MHz band with a single sub-carrier is simulated for the location-based radio map. Measurements from $500$ known locations are used to construct the radio map, as illustrated in Fig.~\ref{fig:interpolartion_location} (right), where the black dots indicate the known locations in the dataset. By comparing the interpolated outage capacities on the right to the true outage capacities on the left, it can be seen that the \gls{gp} successfully manages to capture the main variations across the cell. The predictive distribution from the statistical radio map is then used to select the rate for each unobserved location on the map, where $\delta$ determines how conservatively the rate is selected. As a baseline scheme, we consider the simple rate selection mechanism that picks the $\epsilon$-outage capacity of the closest point in the dataset, i.e., without interpolation. Fig.~\ref{fig:performance_location} shows the empirical \glspl{cdf} of the resulting outage probabilities for both rate selection approaches. We see that the approach based on statistical radio maps manages to meet the desired target reliability at a much higher fraction of time than the baseline, which is due to its ability to take prediction uncertainty into account. In particular, only $0.2\%$ of the users of the proposed method violate the target reliability, whereas the number is $50.1\%$ for the baseline users. Note that selecting the rate conservatively is likely to diminish the average throughput of the system --- see \cite{Kallehauge2022_globecom} for results on this. 

\subsection{Chart-based rate selection}
To illustrate the potential of rate selection based on a channel chart, we consider a dual-band cell in which the \gls{bs} has both a massive MIMO array of $32$ antennas (with two polarizations) and an $800$~MHz MIMO array containing  $4$ antennas with two polarizations. The massive MIMO array operates at $3.5$~GHz, and the \gls{csi} samples are assumed to occupy a total of 288 subcarriers spanning $5$~MHz. The $800$~MHz array operates on $120$ subcarriers spanning $2$~MHz. The training dataset for the chart comprises  $5000$ users, each with a single \gls{csi} sample in the $3.5$~GHz band and several samples of the effective received power in the $800$~MHz band (after applying maximum ratio combining).  
The dimensionality-reducing map used to construct the channel chart is implemented by a dense feed-forward neural network similar to that in~\cite{Ferrand2021}, which takes \gls{csi} samples from the $3.5$~MHz band as input. The network is then trained using a triplet loss function by using the Wasserstein distance \cite[p. 453]{everitt2010cambridge} (i.e., the distance measure) between the rate distributions for each location obtained from the received power samples in the lower band. This promotes that points that are close in the latent space also have similar rate distributions, as seen with the outage capacity illustrated in the upper right plot of Fig. \ref{fig:diagram_map_construction}.

\begin{figure}
    \centering
    \includegraphics[width = .5\linewidth]{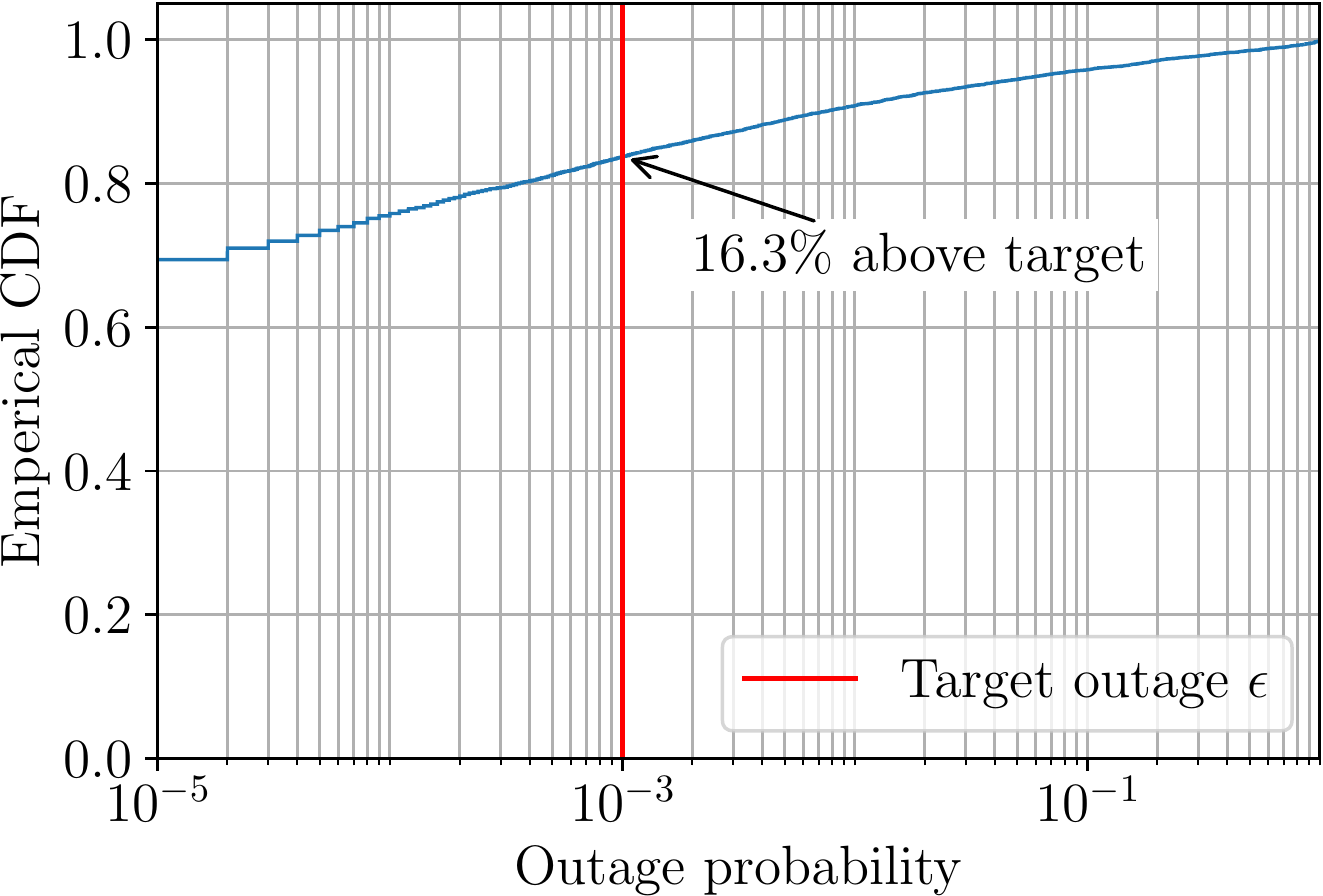}
   \caption{Performance of rate selection using the chart-based radio map with $\epsilon = 10^{-3}$ and $\delta=10^{-3}$. The outage probabilities show that $16.3\%$ of the users violate the reliability requirement.}
    \label{fig:performance_ch_chart}
\end{figure}

The rate is selected by mapping a single \gls{csi} sample from a new user into the fitted channel chart and then computing the quantile of the predictive distribution as in the location-based approach. The resulting distribution of outage probabilities is shown in Fig.~\ref{fig:performance_ch_chart}. 
As with the location-based map, the approach based on the channel chart is able to select a rate that satisfies the reliability requirement with high probability. Although the violation probability of $16.3\%$ is larger than in the case where location information was used, it is still lower than the location-based baseline, which violated the reliability requirement for more than $50\%$ of the users. The reason that chart-based rate selection exceeds the target outage probability more often, is the fact that the predictive variance of the \gls{gp}, which determines how conservative the rate is selected, is much lower compared to the location-based approach. Thus, the \gls{gp} for the channel chart is generally too confident in its predictions, whereas, in reality, there is a significant level of uncertainty in the outage capacity at any given position in the channel chart, as also suggested by the upper right plot in Fig. \ref{fig:diagram_map_construction}.

\section{Future Research Challenges} \label{sec:challenges}
The results from the case study show great potential for leveraging previous measurements in a wireless network by exploiting spatial correlation of channel statistics either in terms of geographic coordinates or coordinates in some latent space, as exemplified by the channel chart. In the following, we discuss some future research directions and potential challenges.

\subsection{Spatial and temporal consistency}
The framework introduced in this article naturally relies on the spatial and temporal consistency of the channel statistics. If the environment varies rapidly in time, the predicted spatial correlation and distributions become obsolete. Note, however, that the changes should be large enough to modify the channel statistics significantly. A procedure to keep the statistical map updated can also be introduced to deal with the temporal dynamics of the channel, for instance, attaching a time stamp to the collected samples and using only the more recent ones to compute the map. 

Regarding spatial consistency, the most favorable case would be a smooth environment where the spatial correlation remains constant over the whole scenario. However, this may be difficult to find, e.g., in urban environments with a mix of narrow streets and open areas. In the use case presented earlier, we have considered the same spatial correlation model for the whole map, but this can be adapted if needed by fitting the model independently over different parts of the map.

\subsection{Channel modeling sample complexity}
The non-parametric approach used to estimate the $\epsilon$-outage capacity based on received power prevents model mismatch, but requires a high number of samples. In fact, more than $1/\epsilon $ samples are needed~\cite{Angjelichinoski2019}, which may become excessive in many practical scenarios, 
especially if the propagation environment is non-stationary so that past observations become outdated after some time. One promising direction in this area is the use of extreme value theory, which provides a parametric distribution for values of a random variable below a certain threshold. Extreme value theory has been applied both in \cite{Angjelichinoski2019} and \cite{Mehrnia2022} to significantly reduce the required number of samples to estimate statistics for the extremely low values of the channel. However, they suffer from being dependent on hyperparameters, which are often difficult to specify.

\subsection{Interference}
In current cellular systems, interference --- either external, inter-cell, or intra-cell --- is usually the limiting factor that determines the received signal quality. For the sake of presentation, we deliberately ignored interference in the previous sections, but accounting for it is undoubtedly crucial in real systems. The interference, like any other statistics, can be captured in a statistical ratio map. However, interference depends on the temporal dynamics of exogenous communication traffic, affected by, e.g., the access methods and resource allocation. This leads to a non-trivial problem in which the configuration of the network itself impacts the statistic that the system tries to predict. 

\subsection{Channel chart distance measures}
The Wasserstein distance, as used in this article, or sampling time, as used in \cite{Ferrand2021}, are just examples of metrics that can be used to train the channel chart, with an impact on prediction accuracy. To the best of our knowledge, no results are available proving the optimality of a specific metric for the purpose of channel charting. 

\section{Conclusion}\label{sec:conclusion}
One of the central challenges in supporting \gls{urllc} is to accurately estimate the outage capacity of the wireless channel, since it either requires a very good channel model or a prohibitively large number of samples. In this paper, we have considered the idea of exploiting spatial correlation of the wireless channel to predict the outage capacity using past channel observations by constructing statistical radio maps that leverages either location or channel charting. We have demonstrated that, by modeling the statistical uncertainty of the radio maps, it is possible to predict the outage capacity conservatively, thereby obtaining an outage probability that is much less likely to violate the target reliability compared to a baseline scheme. Our results show that the location-based approach performs better than the one based on channel charting, but the chart-based approach has the practical advantage of not relying on location information, while still performing significantly better than the baseline. Finally, we have discussed some of the current research challenges that need to be addressed in order to fully be able to employ statistical radio maps to support \gls{urllc} and other wireless services.

\bibliographystyle{IEEEtran}
\bibliography{references}

\end{document}